\documentclass[11pt]{article}
\usepackage[utf8]{inputenc}
\usepackage{lineno}
\usepackage{times}
\usepackage{latexsym}
\usepackage{amssymb}
\usepackage{graphicx}
\usepackage{url}
\usepackage{float}
\usepackage{caption}
\usepackage{subcaption}
\usepackage{array,tabularx,booktabs,multirow}
\usepackage[section]{placeins}
\usepackage{hyperref}
\newcommand{\footremember}[2]{%
    \footnote{#2}
    \newcounter{#1}
    \setcounter{#1}{\value{footnote}}%
}
\newcommand{\footrecall}[1]{%
    \footnotemark[\value{#1}]%
} 

\title{Shared E-scooters: Business, Pleasure, or Transit?}

\author{%
  William Espinoza
\footremember{gt}{H. Milton Stewart School of Industrial and Systems Engineering,Georgia Institute of Technology, 755 Ferst Drive, NW, Atlanta, GA 30332. Email: pvh@isye.gatech.edu}
\and
Matthew Howard\footnote{University of Michigan, Ann Arbor, MI 48109.}
\and
Julia Lane\footrecall{gt}
\and
Pascal Van Hentenryck\footrecall{gt}
}

\begin{document}
\maketitle

\section{Abstract}

Shared e-scooters have become a familiar sight in many cities around the world. Yet the role they play in the mobility space is still poorly understood. This paper presents a study of the use of Bird e-scooters in the city of Atlanta. Starting with raw data which contains the location of available Birds over time, the study identifies trips and leverages the Google {\it Places} API to associate each trip origin and destination with a Point of Interest (POI). The resulting trip data is then used to understand the role of e-scooters in mobility
by clustering trips using 10 collections of POIs, including business, food and recreation, parking, transit, health, and residential. The trips between these POI clusters reveal some surprising, albeit sensible, findings about the role of e-scooters in mobility, as well as the time of the day where they are most popular.

\hfill\break%
\noindent\textit{Keywords}: Shared E-scooters, Mobility, Origin-Destination Pairs, Point of Interest, Time of day.
\newpage


\section{Introduction}

E-scooters have become a familiar sight in cities around the world. When they are not seen traveling on streets, bicycle lanes, and around campuses, they can be found parked at transit stations, in front of stores, restaurants, and apartment buildings, or next to parks, stadium, and parking lots. 

Despite their growing popularity, the role of e-scooters in mobility remains poorly understood \cite{StuttgartScooters}. Are they used as a complement to transit in order to address its first/last problem? Are e-scooters mostly convenience vehicles whose trips replace a short walk or are they part of daily mobility activities of a segment of the population? These are important questions to address as e-scooters have the potential to change the mobility landscape in cities but may also require investment in infrastructures and regulations in order to ensure the safety of their riders \cite{BestPractice}.

This paper is an attempt to answer some of these questions, using a data-driven approach. Starting with raw data describing where e-scooters are located when they are idle, the study first identifies e-scooter trips, i.e., origin-destination pairs over time. The trips are then filtered to remove the substantial noise present in the dataset and, in particular, a large number of short trips (less than a few meters) that do not represent actual e-scooter use. The origin and destination of e-scooter trips are then mapped onto a Point of Interest (POI), obtained using the {\it Google Places API} and, in particular, its {\it Nearby Search} and {\it Text Search} functionalities. Google Places provides a number of predefined types for POIs, which this study augments to capture sites that are potentially important for mobility. Once each trip is associated with two POIs, it becomes possible to group POIs in meaningful categories and to try isolating the purpose of each trip. These trip purposes offer unique insights of how e-scooters are currently being used and what their roles could be in the future of mobility. Although the data-driven methodology presented in this paper is general, the paper focuses on the use of Bird e-scooters in the city of Atlanta and, more specifically, its midtown and downtown sections and their neighboring neighborhoods which capture about 70\% of the ridership. 

The trip purpose analysis is rather illuminating, and somewhat surprising: It reveals that, at this point, {\it e-scooters seem to fill specific mobility needs that complement existing modes}, by offering a convenient and affordable transportation option, at least for some population segments. The analysis also reveals the importance of {\em the time of day} in e-scooter use. The main contributions of the paper can be summarized as follows:
\begin{enumerate}
    \item It presents a data-driven methodology for inferring the purposes of e-scooter trips, starting from raw data on idle e-scooter locations and the Google Places API for identifying POIs.
    \item It provides an analysis of trip purposes for e-scooters in the city of Atlanta, grouping trips along 10 broad categories that include public transit, business, recreation, food, residential locations, and health facilities. 
    \item It analyzes the main trip categories in more depth, diving into more detailed trip purposes that reveal some interesting insights on how e-scooters are currently used. 
\end{enumerate}

The rest of the paper is structured as follows. It starts with a brief of background on the operations of {\it Bird}, a ride-sharing company for e-scooters and a review the existing literature on e-scooters. The paper then describes the dataset used for the analysis and the methodology adopted for identifying meaningful trip purposes. The methodology involves the transformation of the raw data into trips, each of which consists of an origin and a destination and their times, the identification of potential POIs using the Google Places API, as well as the association of trips with two POIs. Each of these steps requires some careful cleaning and calibration, which are documented in each of the sections, given the distance between the raw data and actual e-scooter trips, as well as the difficulty in associating POIs to trips. The trip purposes are then presented in matrix form, where the rows and columns represent broad categories of POIs. Deep dives on the most popular categories, and on the time of use, are also presented. The paper then concludes with implications for mobility.

\section{Background}

Bird \cite{Bird} is a ride-sharing company whose business model emulates bike- and car-sharing companies. A rider can activate a bird e-scooter (a Bird) whenever they see one, drive it to their destination, and drop it anywhere. The cost is \$1 to start a trip and 29 cents per minute. Birds can be operated from 7 A.M. to 9 P.M. Every night after hours, Birds are picked up and taken to charge. The following morning they are dropped off by 7 A.M.

\section{Literature Review}

As mentioned in the introduction, the role of e-scooters in mobility remains poorly understood and has not been studied in depth. The study in \cite{StuttgartScooters} is the closest related work: It concerns the analysis of e-scooter riders in the city of Stuttgart in Germany using various clusterings of the population. Starting with a dataset which contains the e-scooter trips, as well as demographic data, they identified four classes of riders: Power Users, Casual Users (Generation X), Casual Users (Generation Y), and One-Time users. The Power Users account for 40\% of the revenue in their dataset but is a rather small group compared to the others. The analysis in this paper is orthogonal in scope and complement those findings nicely: It is not customer-centric, since no such data was available. Instead, this paper focuses on trip purposes and the role of e-scooters in the mobility landscape. In addition, the methodology followed in our study is fundamentally different: It relies on POIs to identify trip purposes. Interestingly, the analysis provided in this paper also reveals, albeit in an indirect way, the main customer groups for e-scooters. A study of best practices in the management of e-scooters is presented in \cite{BestPractice}. It studies the correlation between e-scooter uses and accidents and provides recommendations on how to improve safety for e-scooter riders. The optimal location of charging stations for e-scooters is studied in \cite{CHEN2018519}. The way riders park their e-scooters in the city of San Jose is studied in \cite{ScooterPark}. E-scooters were also a focus in the study of measuring equitable access to mobility described in \cite{ScooterEquity}. The study reports where e-scooters are available and concludes that e-scooter availability is aligned with vehicle availability and are not evenly distributed or used in low-income neighborhoods. These observations are consistent with the trip purposes identified in this paper.

\section{The Dataset}

Bird stores data for all of their e-scooters and had this information publicly available for a short time. This data was available for cities such as Atlanta, Miami, Los Angeles, Portland, and Charlotte. However, for the purposes of this study, only trips in Atlanta were considered. During the data collection, the Bird's server updates approximately every 10 minutes and stores the time of the last update, as well as the data for all Birds in the city. Each Bird is associated with a unique e-scooter ID and has latitude and longitude coordinates for each system update. However, if the Bird is in use at the time of the system update, the e-scooter ID will not be present on the server and instead will have 'null' in its place.  Additionally, the Bird's server does not store historical data, so the extraction program had to pull the data in real time. This data was conducted using Bird's API over the following dates in 2019: January 26 -- February 1, February 2--5, 10--13 and 15--19, and February 26 -- March 5.

\section{The Methodology}

The goal of this project is to understand how  e-scooters are being used and, in particular, to identify the purpose of every e-scooter trip. To achieve this objective, the analysis of the Bird dataset requires a number of steps that can be summarized as follows:
\begin{enumerate}
    \item Extraction of e-scooter trips from the dataset;
    \item Extraction of Points of Interest (POI for short) using the Google Places API;
    \item Association of POIs to the origin and destination of each trip;
    \item Determination of the purpose of each trip;
    \item Aggregation of trip purposes by POI categories and presentation in matrix form.
\end{enumerate}
Many of these steps are complicated by various factors that are highlighted in subsequent sections.

\section{Trip Identification}

This section reviews the first step of the methodology, trip identification, which includes both extraction and cleaning.

\subsection{Trip Extraction}

The first step consists in extracting trips from e-scooter traces. Each e-scooter is tracked independently on a daily basis. The data provides successive coordinate locations and time-stamps for each e-scooter. These start and end locations, and their times, are then used as the origin and destination of the trip. As a result, each obtained trip has an origin, destination, start and end times (month/day/year/time), and a distance of displacement. There are 2.6 million data points in the dataset which spans the days mentioned earlier.

Due to the nature of Bird's server, which only updates once every 10 minutes, these trips are necessarily an approximation. First observe that, if a trip starts after a given time-stamp $t_1$ and ends before the system updates again at $t_2$, the trip has a start time of $t_1$ regardless of how close the actual start time was to $t_2$. Similarly, any ride that ends after a given time-stamp (say $t_3$) but before the system updates again (at $t_4$) produces a trip with an end time of $t_4$. As a result, this definition of trip cannot capture a situation where two trips start and end in between two time-stamps. 

\subsection{Trip Cleaning}

\begin{figure}[!t]
   \centering
    \begin{subfigure}[b]{0.48\textwidth}
        \centering
        \includegraphics[height=2.0in]{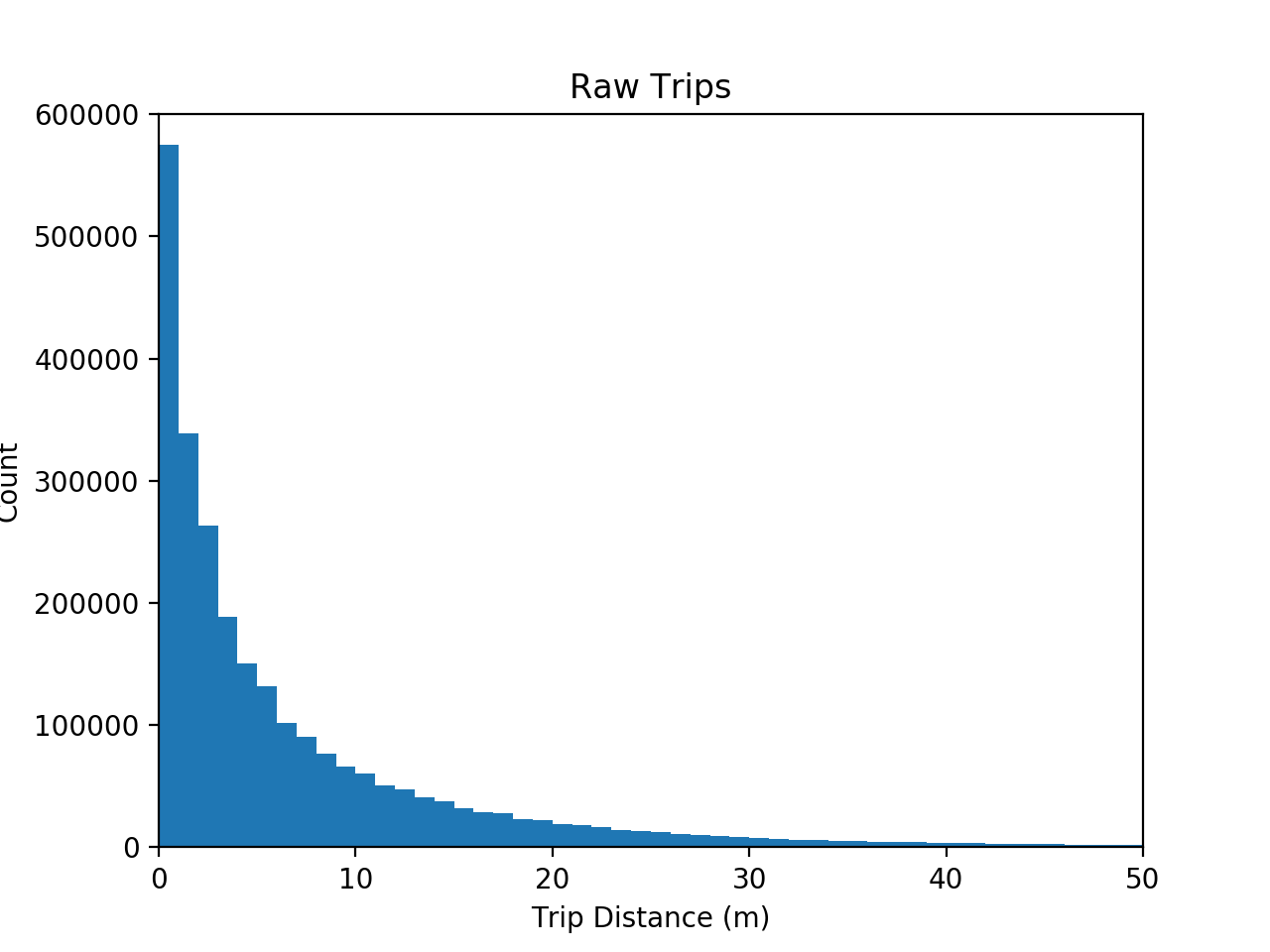}
        \caption{The Raw Trips.}
    \end{subfigure}%
    \begin{subfigure}[b]{0.48\textwidth}
        \centering
        \includegraphics[height=2.0in]{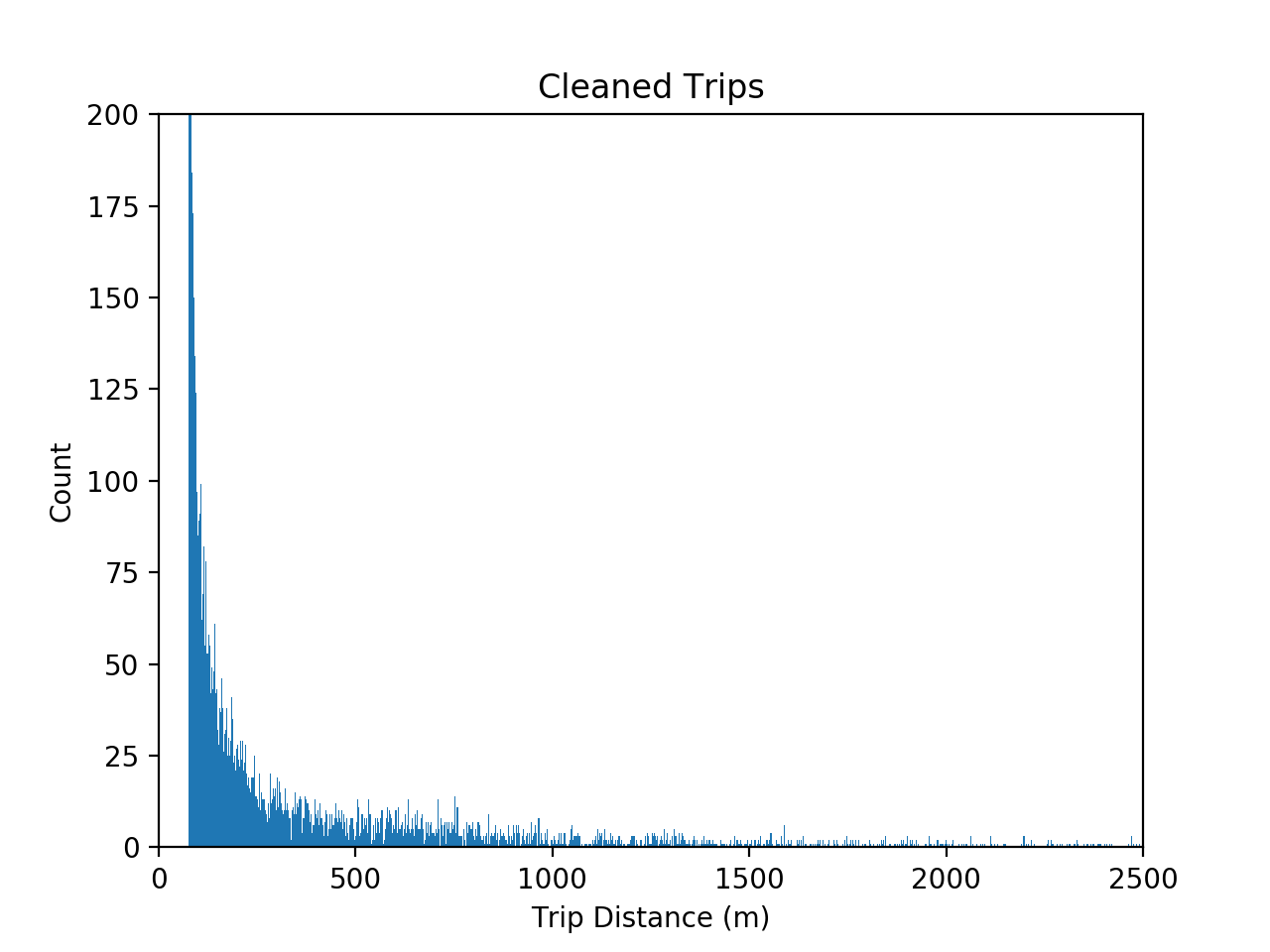}
        \caption{The Cleaned Trips.}
    \end{subfigure}%
    \caption{Illustrating the Trip Cleaning Process.}
    \label{fig:trips}
\end{figure}

The raw trips then need to be cleaned. First, because of Bird's operating hours, any trip that has a start time or end time outside of the 7am to 9pm range is excluded. This decreases the chance of classifying as trip when an e-scooter being picked up for charging. All trips under 75 meters and over 3000 meters are also eliminated. This is  done in part to eliminate any charging trip that occurs during the day, as well as  short distance trips that are assumed to be noise. Surprisingly, in the dataset, there are 1,516,631 rides under 5 meters, 1,983,723 under 10 meters, and 2,354,293 under 20 meters, which provides compelling evidence that these rides do not correspond to actual trips. After this filtering steps, there are around 23,000 trips left, which is much more reasonable.
Figure \ref{fig:trips} highlights the raw and cleaned trips.

\section{Points of Interest}

This section describes the methodology to obtain the POIs from the Google Places API.

\subsection{Google Places}

This study uses {\it Google Places} \cite{GooglePlaces} to obtain POIs: 
{\em Places} has various capabilities and this study employs two of them: the {\it Nearby Search} and the {\it Text Search}. Both of these conduct a search within a given radius of a specific coordinate. {\it Nearby Search} relies on {\bf predefined types} provided by Google in \cite{GooglePlacesPredefined}: Hence {\it Nearby Search} has the benefit to pull results that match the specific predefined type specified by queries. The {\it Text Search} pulls information on the POIs that match the text string in the query input: It has a slightly lower degree of reliability.

To keep the Google search budget reasonable, the city is divided into a grid. Once the city is embedded in this grid, it quickly becomes apparent that one grid cell is responsible for 70\% of the total traffic: That cell captured the (important) midtown and downtown areas. As a result, only trips in this region are analyzed. The upper right and lower left coordinate boundaries for this specific grid are [33.789279,-84.35961499999999] and [33.74837933333333,-84.40562333333332]. This leaves a final count of 16,217 total trips.

Although Google Places is powerful, it is not without limitations. In order to prevent mass data extraction by each query, Google does not return more than 20 results. To overcome this, an even smaller grid is created out of the new coordinate boundaries (8x8, 64 total grid points). For each grid point, a search is conducted using the center of the grid point and a calculated radius that would ensure the grid point is circumscribed within the query area. 

\subsection{General Points of Interest}

Out of the 90 predefined types for {\it Nearby Search}, only those deemed relevant for the purpose of this study are used in order to allow for as many queries as possible for a given query budget. There are, however, gaps in predefined types for {\it Nearby Search}. More precisely, there is no residential type. POIs such as apartments and condos are thus obtained using the {\it Text Search}. Note also that the {\it Nearby Search} associates with each POI its name, location, predefined types, and vicinity (address). In general, there are about 2--3 predefined types returned for a given POI.

\subsection{Additional Points of Interest}

To increase the overall POI density in the selected region, some additional procedures are conducted. After reducing the selected region to a 15x15 grid, more queries are run for the types "condo", "lodging", "park", "restaurant", and "subway station" in the grid points with the lowest POI density. Thirteen business points of interest, corresponding to major corporations, were also manually created to overcome spurious parking associations. Lastly, three neighborhoods were manually added, one being in Midtown, one being Old Fourth Ward, and one being Virginia Highlands.

\subsection{Primary Types}

For classification purposes, this study also associates with each POI a {\em primary} type. The set of primary types is composed of Google's predefined types and the types introduced by this study (e.g., apartment). The primary type of a POI obtained by the {\it Nearby Search} is the first predefined type returned by Google. This selection is based on the assumption that Google returns the list of associated types in decreasing order of relevance.
 For those added through the {\it Text Search} whose respective types do not exist in the Google API, their primary type was programmatically added to their list of associated types. 
 
 \subsection{POI Buffers}
 
There are some limitations with using a specific coordinate point. Consider the Mercedes-Benz stadium in Atlanta. Because its coordinates are in its center, the stadium has a low probability to be associated with Birds, which are not allowed inside the building. As a result, a restaurant or coffee shop with a more readily available coordinate could end up being associated with a Bird that was parked right next to the Benz. This concept can be extended to other points of interest such as  MARTA stations, the Georgia Aquarium, and neighborhoods. To address this limitation, a series of "buffer" points of interest are programmatically added around these key POIs using an azimuthal equidistant projection. 

\begin{figure}[!t]
   \centering
    \begin{subfigure}[b]{0.48\textwidth}
        \centering
        \includegraphics[height=2.0in]{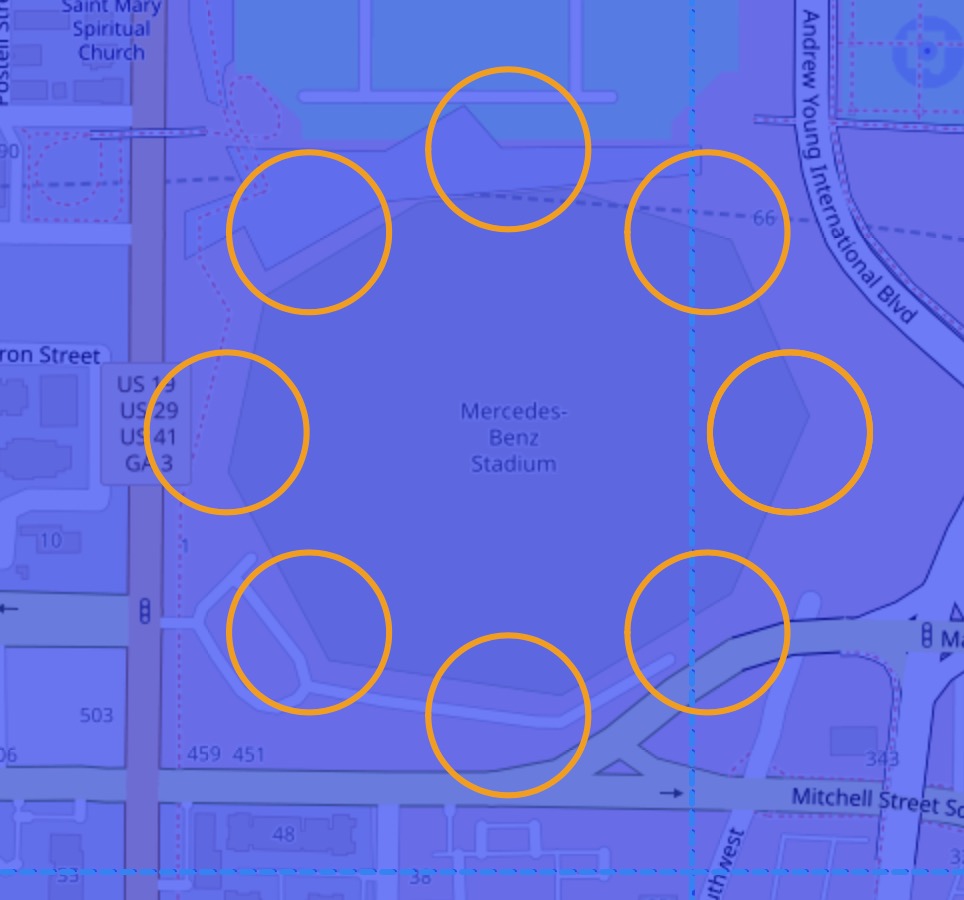}
        \caption{The Mercedes-Benz Stadium.}
    \end{subfigure}%
    \begin{subfigure}[b]{0.48\textwidth}
        \centering
        \includegraphics[height=2.0in]{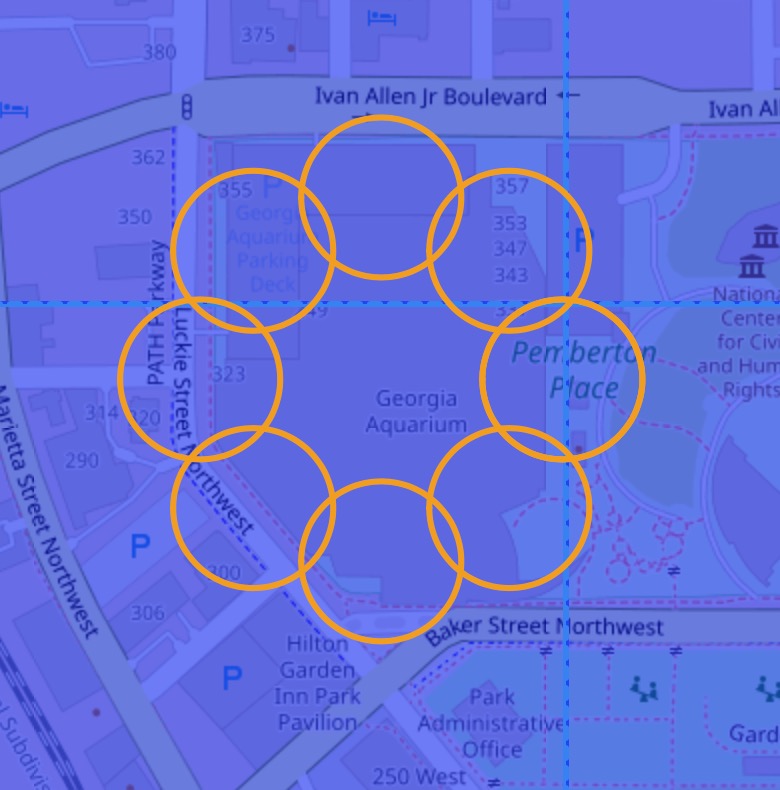}
        \caption{The Georgia Aquarium.}
    \end{subfigure}%
    \caption{Illustrating the POI Buffers.}
    \label{fig:buffers}
\end{figure}

Eight buffers are given to the Georgia Aquarium (radius = 90 m), Mercedes-Benz Dome (radius = 140 m), MARTA stations (radius = 20 m), and neighborhoods (radius = 140 m). To reduce the distance between the POIs associated with a trip within a neighborhood, an additional ring of 16 buffers are added around all 3 neighborhoods (radius = 290 m). These buffer points inherit the same data as their parent POIs. POI buffers are illustrated in Figure \ref{fig:buffers}.

\subsection{Duplicate Removal}

After collecting all POIs, it is necessary to address duplicates entries as well as other issues. Indeed, a search on "restaurant" and a search on "bar" may return the same POI. In this case, the {\it Nearby Search} result whose primary type does not match the query type is removed. More generally, in presence of duplicates, only the POIs whose primary type matches their first predefined type are kept. Of course, none of the results from the {\it Text Search} can be post-processed in this way since these queries are not based on predefined types.

Another complication comes from the fact that there are many POIs with only Atlanta in their vicinity field as opposed to a proper address. Upon further investigation, none of these POIs are assigned proper coordinates and it appears Google does not have an accurate location information on them. All of these are also excluded. 

The last change to the POI data is motivated by the fact that different POIs may share the same exact location. As a result, it is not possible to associate a trip purpose to an O-D pair with that location. To capture this uncertainty, a new primary type is created and called {\em multiple}. All POIs sharing a location are then replaced by a single {\em multiple} POI. The new instance inherits all of the primary types, including duplicates, from its predecessors. Note that, when the two POIs are of the same primary type, the merging process still occurs but the primary type is kept.

\subsection{Grouping POIs}

\begin{table}[!t]
    \centering
    \includegraphics[height=2.5in]{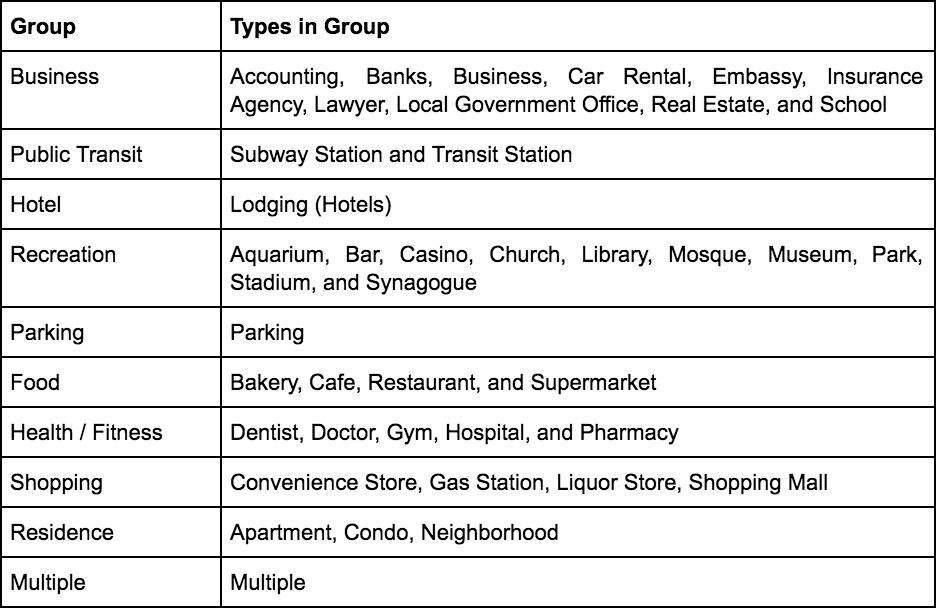}  
    \caption{The 10 POI Groups Used in This Study.}
    \label{table:group}
\end{table}

Because this study relies on 42 primary types of POIs, these are further categorized into 10 different groups described in Table \ref{table:group}. Bus stops could also be included, but this would result in an extremely high volume of associations, most of which are deemed to be unreasonable. Future work should cross-analyze Automatic Passenger Count and GPFS data with bus stops to determine potential associations to bus stops. As a result, after this step, every POI is thus endowed with a group and a primary type.

\section{Linking Trips and POIs}

With clean trips and POIs, it becomes possible to associate a POI with the origin and destination of a trip. For each trip coordinate (i.e., origin or destination), the closest POI is found using a k-dimensional tree. Because {\tt scipy}'s k-dimensional tree uses euclidean distances, the POIs, origins, and destinations are first converted to a Cartesian plane. The trips are then augmented with their origin and destination POIs. Moreover, the distances between the origin and destination and their POIs are also added. Indeed, these distances help in establishing the likelihood of the inferred trip purpose. In the case where the same POI is returned for both origin and destination, an additional Google query is run  on the origin location. The choice of the origin for reassignment is motivated by the fact that a Bird rider has more control over her destination than her origin.

\subsection{Distance Threshold Sensitivity Analysis}

\begin{figure}[!t]
   \centering
    \begin{subfigure}[b]{0.48\textwidth}
        \centering
        \includegraphics[height=2.3in]{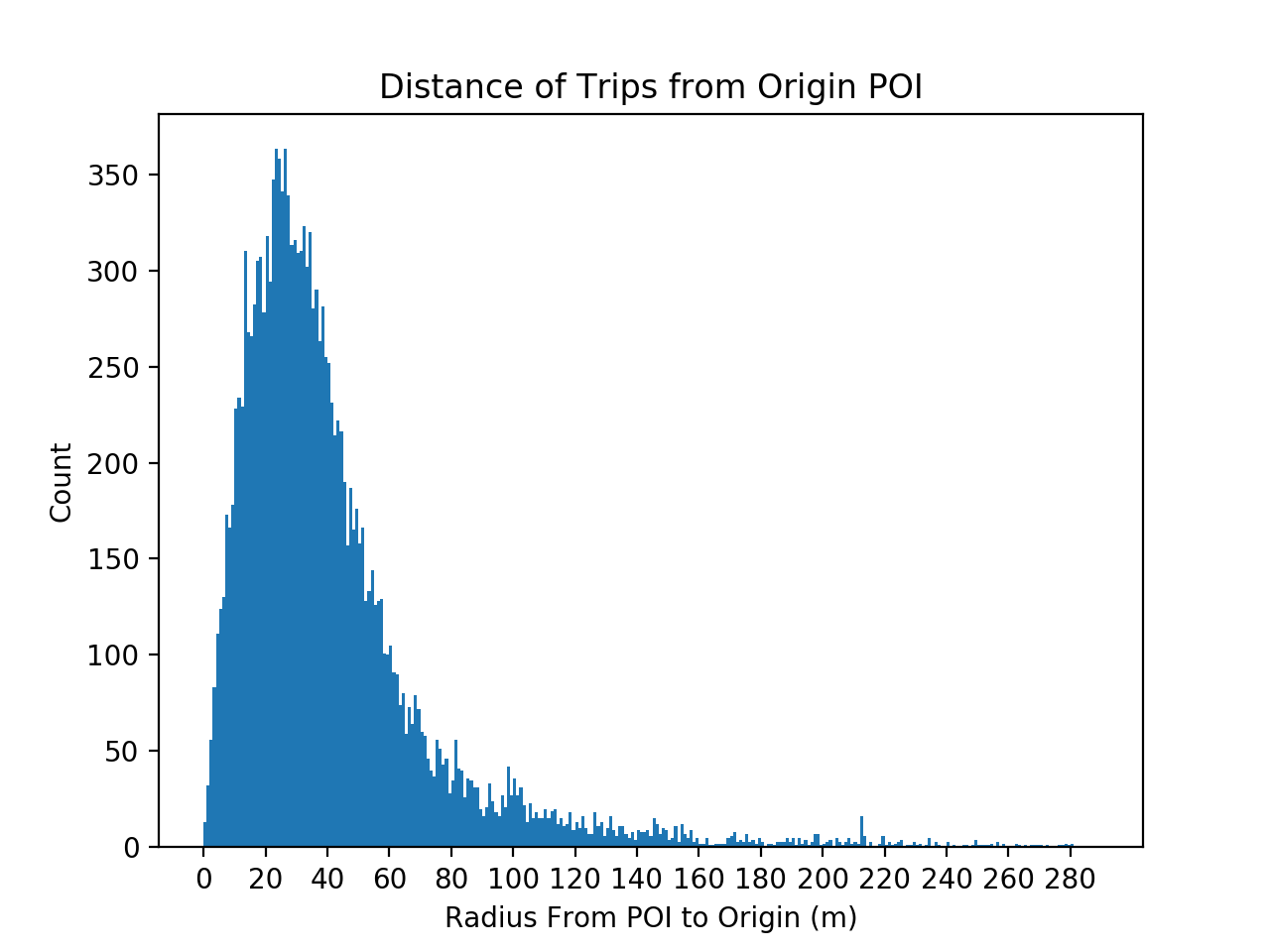}
        \caption{Distances between Origins and POIs.}
    \end{subfigure}%
    \begin{subfigure}[b]{0.48\textwidth}
        \centering
        \includegraphics[height=2.3in]{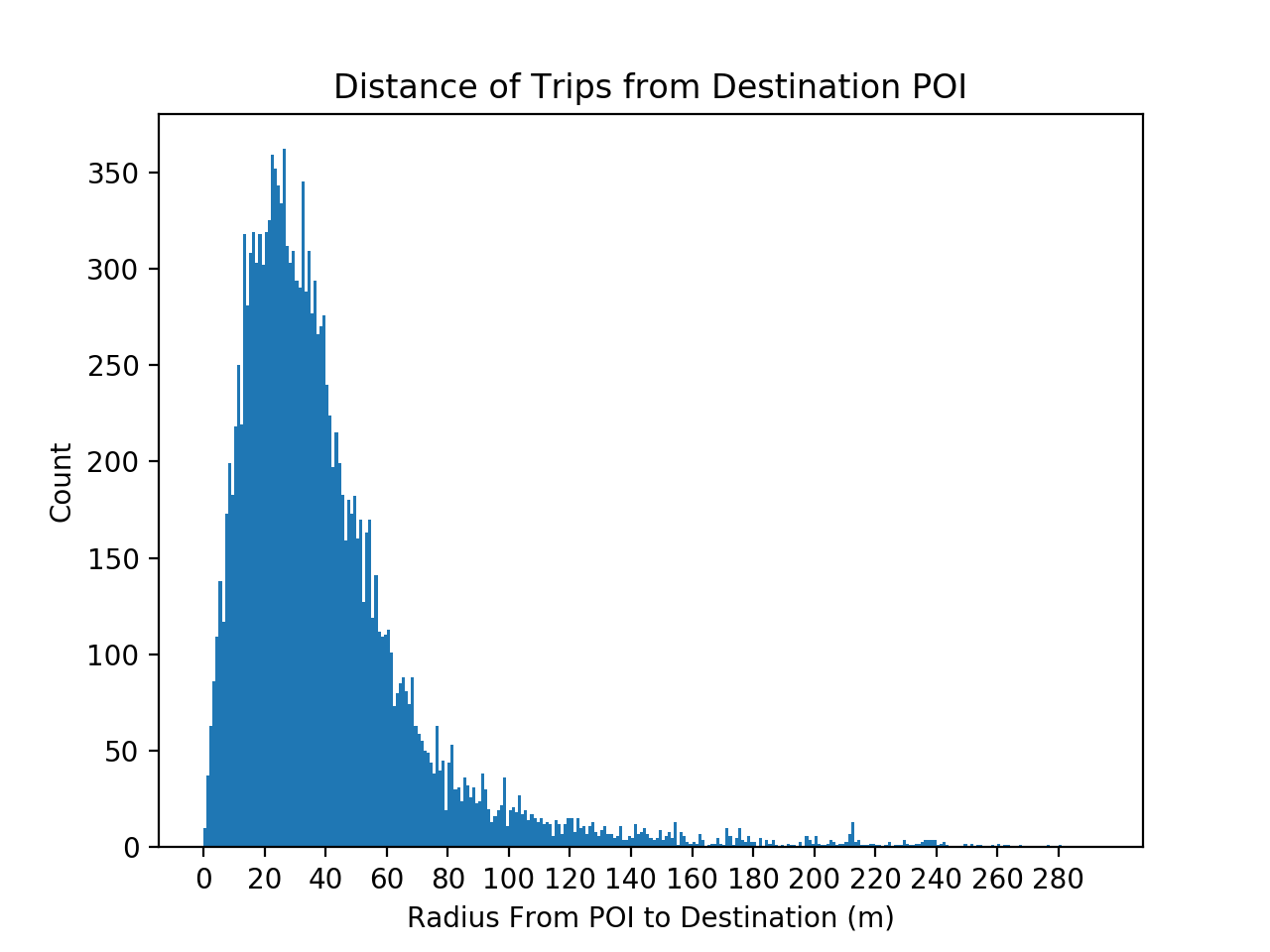}
        \caption{Distances between Destinations and POIs.}
    \end{subfigure}%
    \caption{Distances between Origins/Destinations and POIs.}
    \label{fig:distances}
\end{figure}

Although the origins and destinations of all trips are assigned a POI, there are significant variations in distances between origins/destinations and POIs, as shown in Figure \ref{fig:distances}. In general, the larger the distance between an origin/destination and its POI, the lower the confidence in the association. This means that, in order to eliminate spurious associations, a certain distance threshold needs to be established. To find such a suitable cutoff point, it is useful to look at how the number of trips increases with distance for various groups of POIs.  

\begin{figure}[!t]
    \centering
    \includegraphics[width=10cm]{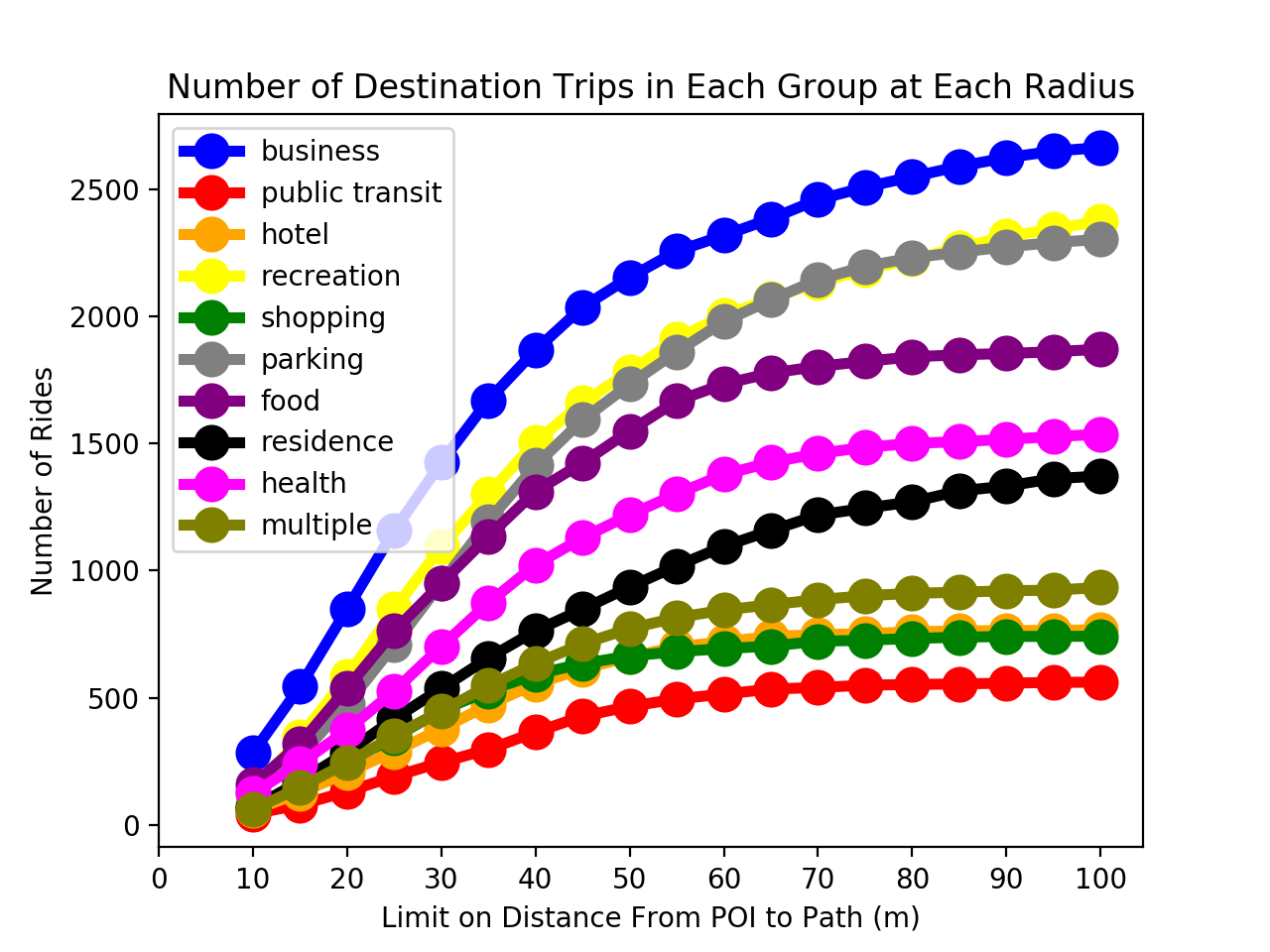}
    \caption{Number of Trip Destinations per POI Group as a Function of POI Distance.}
    \label{fig:POIgroups}
\end{figure}

\begin{figure}[!t]
    \centering
    \includegraphics[width=10cm]{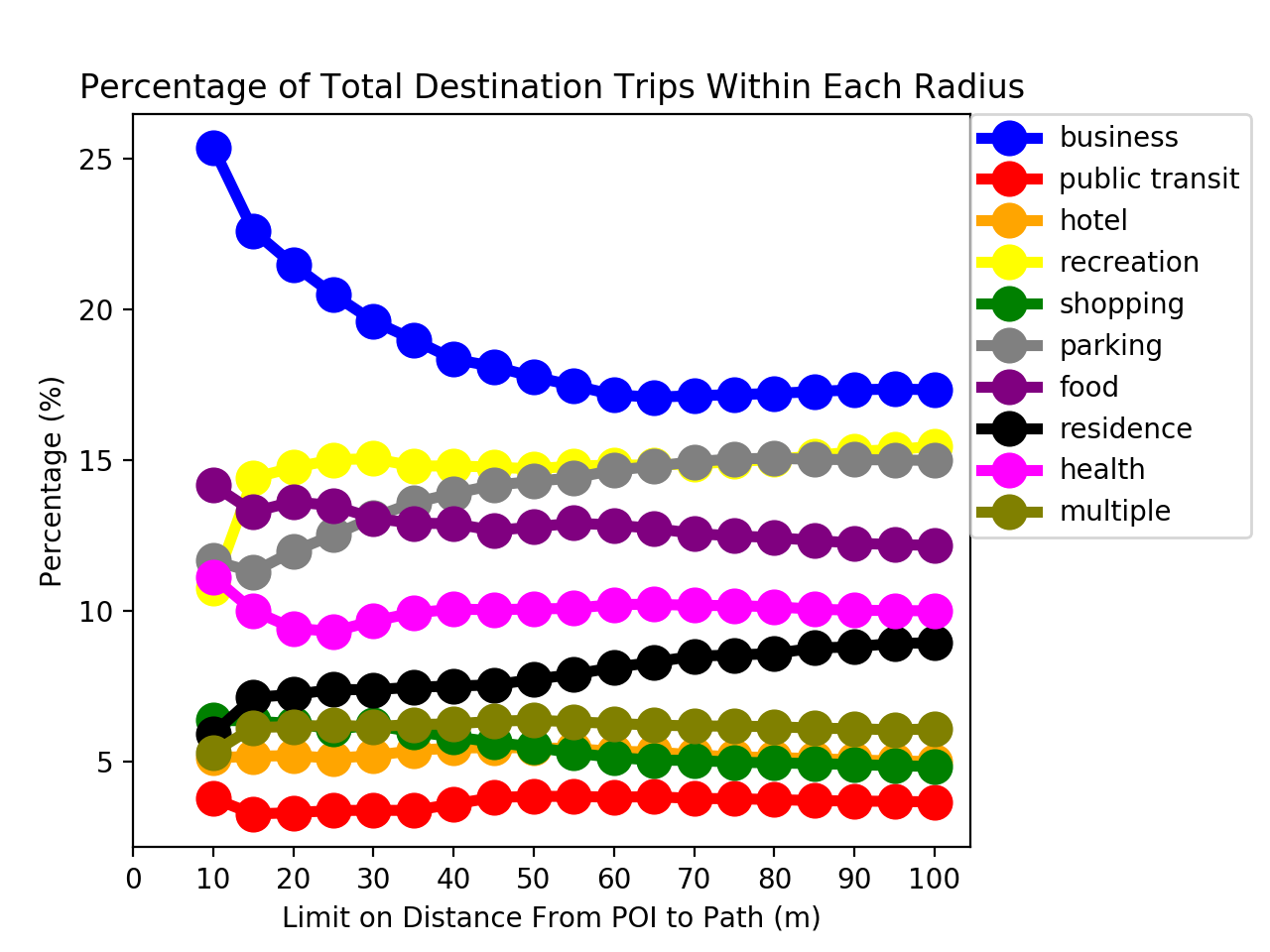}
    \caption{Percentage of Trip Destinations per POI Groups as a Function of POI Distance.}
    \label{fig:POIgroupsPercent}
\end{figure}

Figure \ref{fig:POIgroups} depicts the results: Across POI groups, the number of trips at a given threshold displays a similar trend of logarithmic growth as the distance increases. This means that the main increase in trip/POI associations occurs for distances within the 0 to 50 meter range. Beyond 50 meters, the increase in  trip/POI associations is minimal and probably not reliable.

The same information expressed in percentage is shown in Figure \ref{fig:POIgroupsPercent}: It tells a slightly different, if not still reaffirming, story. The percentages of trips associated with each POI group outside of {\it business}, {\it parking}, and {\it residence} have minor fluctuations. {\it Business} shows a sharp decrease as the distance threshold increases, and {\it parking} and {\it residential} POIs compensate for its decrease. As all groups find a relatively stable percentage at 50 meters, this is determined to be most suitable threshold going forward. Note that the percentage decrease in the POI group {\it Business} indicates that business associations correspond to small distances and hence should be highly reliable. As will become clear, this is reassuring given the conclusions of this study. 

As e-scooters provide a high amount of flexibility for transportation, 50 meters may seem to be quite far. However, this is not necessarily the distance from origin/destination to the nearest POI. In reality, the coordinates returned from Google are generally in the center of a point of interest. A rider, therefore, would drop off her e-scooter near an entry point to the location, rather than the exact coordinate in the middle of the building. When taking into account, a 50 meter threshold becomes very reasonable. In total, out of the 16,217 paths, 9,531 have both an origin and destination associated at 50 meters or less.

\section{Trip Purposes}

It is now possible to try to understand trip purpose for Birds. 

\subsection{Overall Trip Purposes}

\begin{table}[!t]
    \centering
    \includegraphics[height=1.7in]{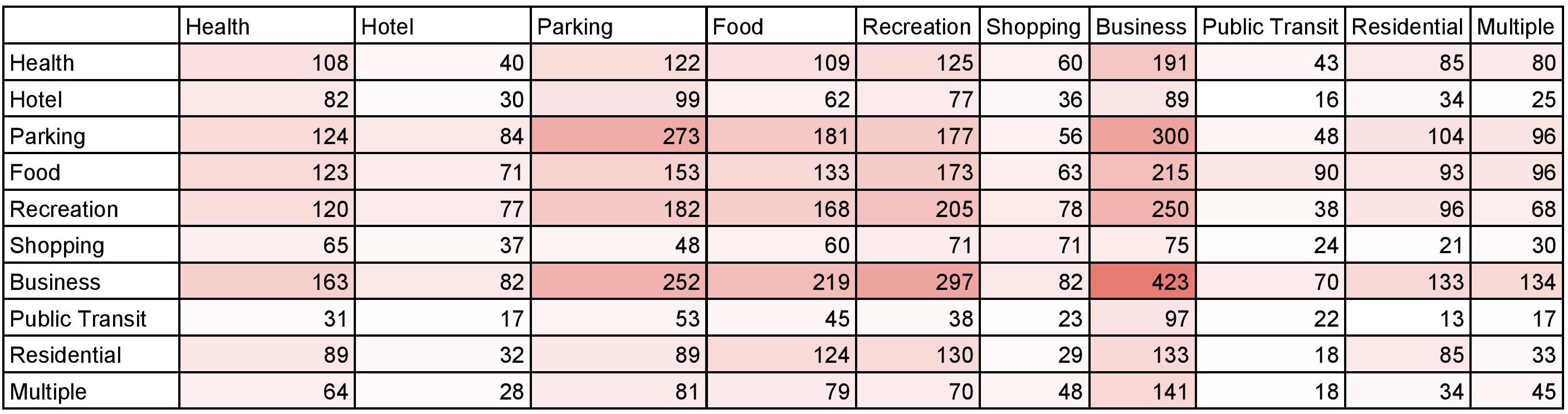}  
    \caption{Trip Purposes: A 2-Dimensional View. The x-axis are the origins and the y-axis are the destinations.}
    \label{table:TP}
\end{table}

Table \ref{table:TP} presents a 2-dimensional view of the trip purposes: Each row considers the trips with an origin specified by its POI group: It reports the number of trips going to each POI group specified by the columns. Observe that the POI groups {\em Parking}, {\em Food}, {\em Recreation}, and {\em Business} consistently stand above the rest. This holds true for both origins and destinations. The rest of this section dives deeper in the information provide by this matrix

\subsubsection{Business to Business}

Surprisingly, {\it Business} to {\it Business} is  the trip purpose with the highest counts. This suggests that Birds are being used as travel to and from business meetings in the city. Birds have a financial advantage over Uber and Lyft and a convenience advantage over MARTA (the Metropolitan Atlanta Rapid Transit Authority), making them easy and affordable to move quickly between businesses. 

\begin{table}[!t]
    \centering
    \includegraphics[height=1.25in]{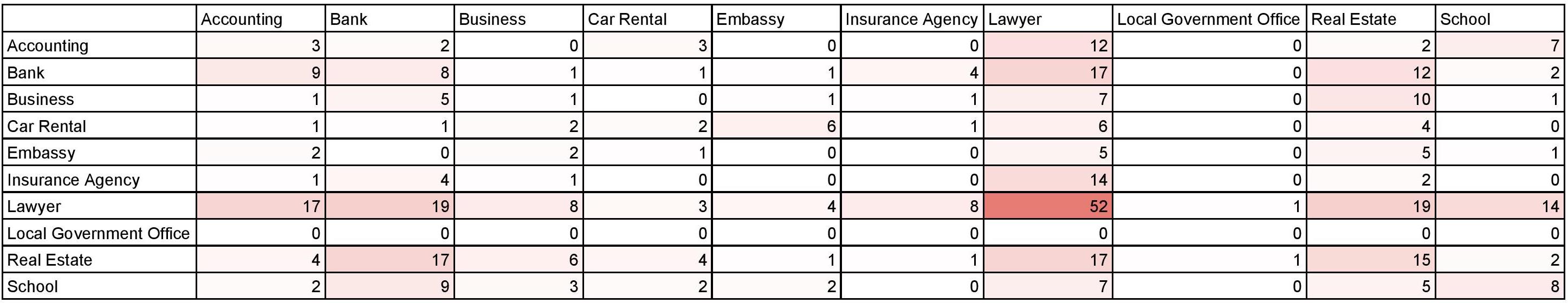}  
    \caption{Zooming into the Business Trips.}
    \label{table:business}
\end{table}

Drilling down further into the business group, out of all of the Business to Business trips, a total of 145 trips are coming from the {\it lawyer} type and a total of 137 are going to a {\it lawyer} type. {\it Real Estate} was the second highest type with 74 and 68 respectively. This is shown in Table \ref{table:business}. What is unique about these jobs is their requirement for a certain degree of mobility. Many lawyers represent a vast array of clients, and this means that they could be visiting all sorts of potential POI types. It could be for a sit-down meeting in a place of business, a casual meeting, or even accompanying a client to court. Real estate agents will frequently leave the office to show a prospective buyer a new property. Ultimately they will have to return back to the agency, and 44 of the residential to business trips have a real estate agency as the associated destination. 

From the opposite perspective, these are occupations that may incur client visitations. This same principle could be applied to the bank type which is responsible for 65 origins and 55 destinations for {\em Business}. A Bird would then provide a flexible mode of transportation for the client. It is then possible that these clients are in a tax bracket where the convenience incentive of a Bird outweighs the financial burden. There is also no waiting time for a Bird if it is available. If it is not, the rider may choose another, more expensive, option. {\em In summary, Birds seem to have addressed a mobility gap in first/last mile business trips, for which they bring  convenience and affordability, at least for some population segments.}

\subsubsection{Business from/to Parking}

This idea can be expanded upon when examining both the high volume of {\em parking} to {\em business} trips and their inverse. In this case, workers are in a situation where they can drive their own car to work. However, no parking lots may be available close to their places of work, and those that are can be quite expensive. A Bird can serve as a way of alleviating the stress and financial burden of finding a parking spot close to work. Instead, a commuter could park further away, and use the Bird to quickly get through the last leg of their trip. {\em In this case, unlike with Public Transportation, Birds are being used as  last-mile solution. If confirmed, this would be encouraging for cities which could relocate parking lots outside the main business areas and use Birds for the last mile.} 

\subsubsection{Leisure}

Moving on, {\em recreation} and {\em food} are two of the other most dominant categories, and even more so in the evening and at night. Out of the 205 recreation to recreation trips, 92 involved leaving a bar and 103 were going to a bar. In addition, there were 89 instances of a trip going from a restaurant to a bar and 78 going from a bar to a restaurant. There are a couple of explanations for this, and they are not necessarily mutually exclusive.

Riders for these trips could be operating under the same incentives as those contributing to the business to business trips. For those individuals who may have to Uber/Lyft into the city or live in the city, using a Bird is a good trade-off between cost and convenience. Indeed, if all of their destinations are in a relatively similar area, the fixed cost of ordering an Uber/Lyft would not be competitive. {\em Again, if confirmed, Birds may have found a sweet spot in mobility for food and recreation trips.}

The second explanation would again incorporate parking into the equation. 84 trips go from a parking lot to a bar and 93 go from a bar to parking. On top of that, 136 are from a restaurant to parking and 145 are from parking to a restaurant. In this situation, a user would be coming into the city with their own car. Although they have an additional trip to the parking lot, the same kind of incentives are at work. Note that the data shows that e-scooters are used in combination to parking to go to bars and restaurant: It is not simply visiting a bar or a restaurant before returning home after a e-scooter trip to parking. {\em Once again, e-scooters seem to address a last-mile need.}

\subsubsection{Transit}

Looking at this trend, it does not appear that Birds are being commonly used as a last-mile complement to MARTA, the Atlanta Public Transportation system. This does not rule out compatibility between the two services in the future but it may suggest that the convenience incentive for MARTA riders to use Birds is outweighed by its financial burden. 

\subsection{Trips Over Time}

\begin{figure}[!t]
   \centering
    \begin{subfigure}[b]{0.48\textwidth}
        \centering
        \includegraphics[height=2.5in]{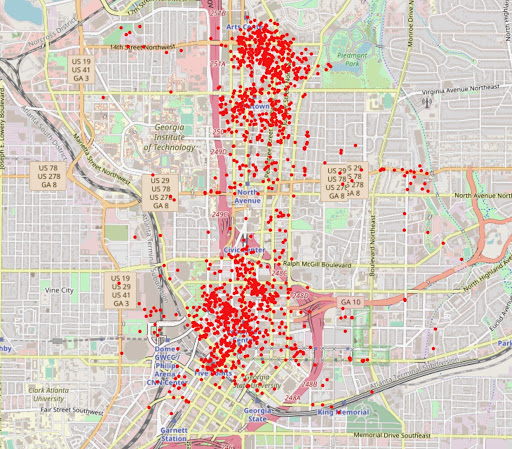}
        \caption{Trip Origins in the Morning.}
    \end{subfigure}%
    \begin{subfigure}[b]{0.48\textwidth}
        \centering
        \includegraphics[height=2.5in]{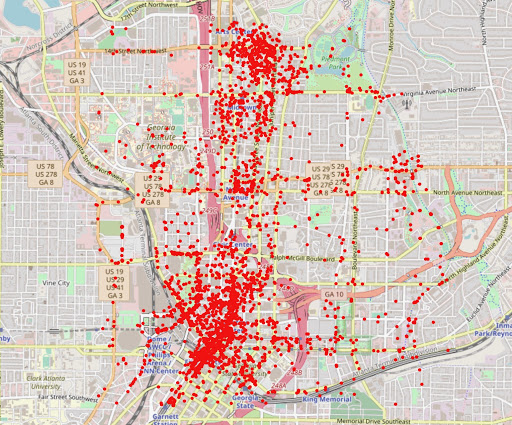}
        \caption{Trip Destinations in the Evening.}
    \end{subfigure}%
    \caption{The Effect of Time of Day on the Number of Trips.}
    \label{fig:timeofday}
\end{figure}

The effect of time of day on the number of trips is profound as shown in Figure \ref{fig:timeofday}. The figure depicts Bird's trip locations in the morning (left) and in the evening (right) respectively. The morning has a much lower density of trips in the city and is also confined to Midtown and Downtown clusters. By the end of the day, both the trip density inside and outside of these two clusters has increased dramatically. There is also a notable amount of trips on major roads and the belt-line in the eastern part of central Atlanta. From examining these images, it is thus not surprising that a significant portion of trips is associated with recreation, business, and food.

\subsection{Trip Purposes By Time of Day}

Since mornings and evenings behave rather differently, this section explores the relationship between trip purposes and time of day. To study this dependency, trips are divided into 5 different time slots; The 7:00 AM - 10:59 AM slot is designed to capture the morning commute; 11:00 AM - 1:59 PM is aimed at capturing the lunch hour;  2:00 - 4:59 PM is used for the afternoon hours, and 5:00 PM - 6:59 PM is used for commuters and restaurant patrons. The last time slot, 7:00 PM - 8:59 PM, is primarily geared towards capturing night life activity before Bird's end of operational hours.

\begin{table}[!t]
    \centering
    \includegraphics[height=2.4in]{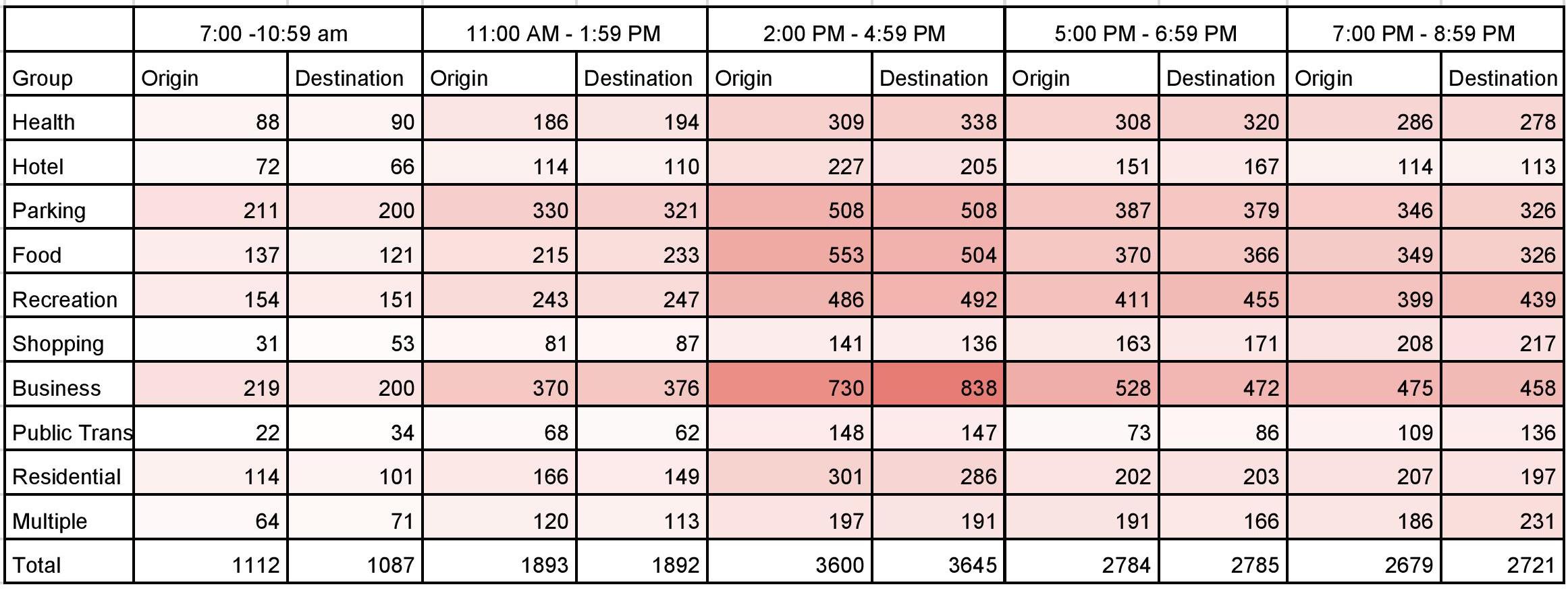}  
    \caption{Trip Purposes by Time of Days.}
    \label{table:TPTOD}
\end{table}

The results are shown in Table \ref{table:TPTOD}. The results indicate that the {\em Parking}, {\em Food}, {\em Recreation}, and {\em Business} groups are relatively similar when comparing the origin and destination for a given time slot. 

The main take away from Table \ref{table:TPTOD} is the magnitude of changes in ridership throughout the day. The ride counts in the morning and at lunch are fairly modest, there is  a large increase in the afternoon, which is generally maintained in the evening and at night. Observe that the afternoon time slot has the benefit of an extra hour when compared to the night and evening slots. In general, the number of rides increases throughout the day and there is an overall lack of symmetry in rides between the early and later hours.

It is also interesting to observe that business travel peaks in the afternoon, which could suggest that Birds are being used as afternoon travel to and from meetings in the city. Similarly, recreation and food are especially prominent in the evening and at night.
Indeed, the high overall volume of trips after hours suggests that Birds are being primarily used as a vehicle of leisure at that point.

These time of the day results further increase the confidence of the POI associations. The fact that recreation and food arise in the evening and at night and business travel in the afternoon is particularly reassuring.

\subsection{Weekdays Versus Weekends}

\begin{table}[!t]
    \centering
    \includegraphics[height=1.35in]{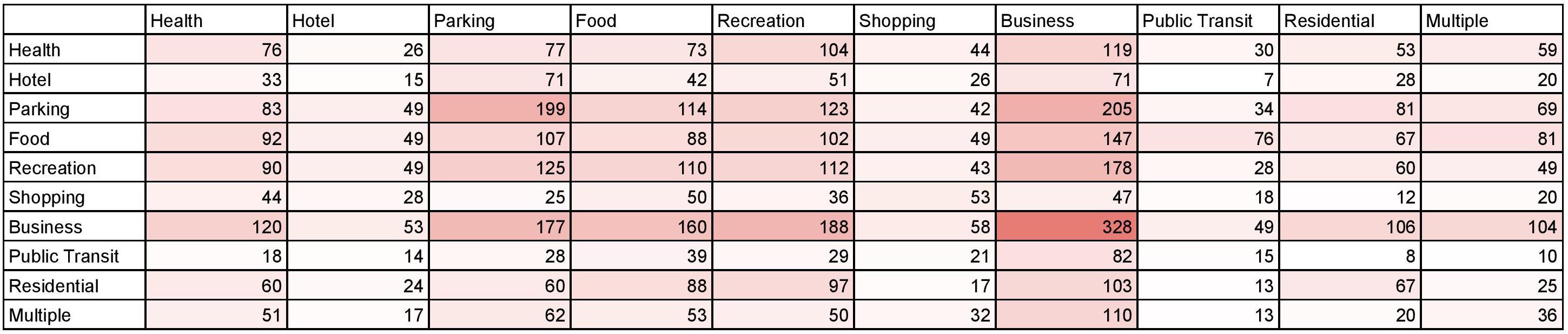}  
    \caption{Trip Purposes on Weekdays.}
    \label{table:TPWD}
\end{table}

\begin{table}[!t]
    \centering
    \includegraphics[height=1.35in]{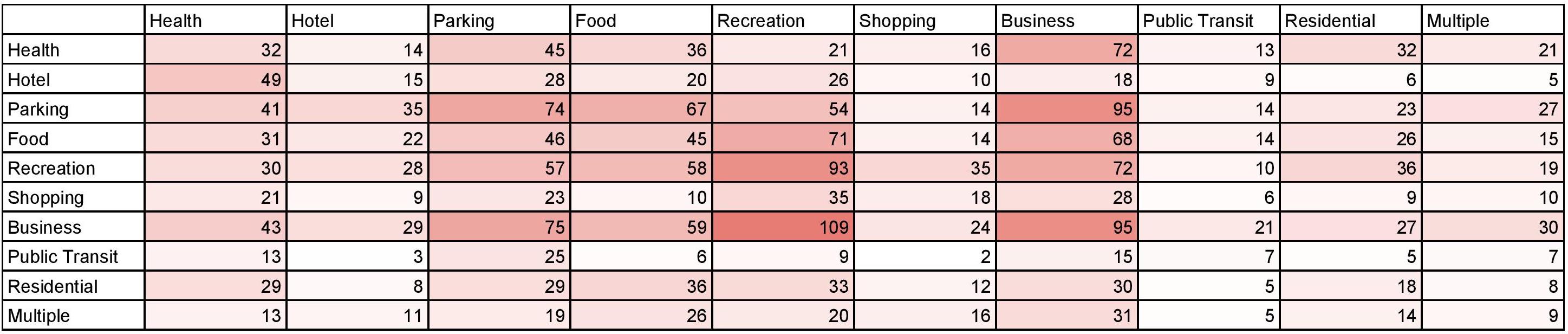}  
    \caption{Trip Purposes on Weekends.}
    \label{table:TPWE}
\end{table}

Tables \ref{table:TPWD} and \ref{table:TPWE} report the trip purposes for weekdays and weekends. The same trends as before can be observed, with {\em Parking}, {\em Food}, {\em Recreation}, and {\em Business} consistently standing above the rest. The main difference between weekdays and weekends is of course the share of business trips which decreases significantly on weekends as expected. It also emphasizes the role of e-scooters for leisure: Food and recreation have significant ridership both on weekdays and on weekends.

\section{Implications}

This analysis indicates that Birds are a highly versatile method of transportation. Not surprisingly, given their ubiquity, they fill a number of gaps in the mobility space. In particular, this paper has shown that Birds are heavily used for business to business trips, business from/to parking, and for recreation. For these POI groups, Birds provide a new option for last-mile mobility. The Business from/to Parking trips are especially interesting for cities: They indicate that Birds may be a technology enabler to relocate parking spaces outside the city center. Birds, together with infrastructure improvements to ensure safety, may thus have interesting consequences for urban planning. Note however that Bird usage currently appears to be limited to a certain demographic. This is generally individuals who can afford an option that is more convenient and flexible than transit/walking and yet remains affordable. However, cities may have incentives to make Birds more accessible. 

It is important to emphasize that there is also a significant time of the day aspect, with the afternoon and evening hours having significant more rides than mornings. To a large extent, this is a consequence of the trip purposes obviously as discussed earlier.

The use of Bird for transit is however relatively small, although Birds are typically available around public transit stations. It is possible that, at its current price, the financial burden of using a Bird as a part of a commute is too high for transit users. Indeed, in Atlanta, a simple Bird trip from the Midtown MARTA station to Georgia Tech is \$4. In contrast, for business trips, there is a demonstrated ability for Birds to fulfill the last mile need. 

A packaged deal between Bird and Transit (MARTA) could open the door to a previously untapped market. In order to gain access to more consistent and reliable usage, Bird could offer a financial incentive to MARTA users. This could come in the form of a free ride every 10 trips or discounted fares. Potentially when users purchase a MARTA pass, for an extra fee, a Bird subscription could be added. 

In order to accomplish a working relationship, there are still two main issues to tackle. From an economic perspective, removing the financial deterrent would require the price to be low enough to satisfy users, but high enough for Bird to have a sustainable business model. After addressing this, the next main issue would be the reliability of a Bird commute. If a user was going to consistently commit to commuting with Bird, they will want Bird to commit to consistently having a e-scooter available for them. This could be accomplished with some sort of reservation feature or by Bird calculating the expected demand for e-scooters at each station. This would cause a few things to change. First, bird could see a consistent, reliable revenue stream unlike before. Second, a whole new group of users now have an affordable and flexible transportation option. Third, MARTA by extension would become a much more flexible and convenient option. Instead of seeing MARTA as limited, this could open the door to a more encompassing public transportation system.

\section{Conclusion}

Shared e-scooters have become a familiar sight in many cities around the world. Yet few studies have analyzed the role they play in the mobility space. This paper originated from an attempt to fill this gap. Starting from raw Bird data for the city of Atlanta, it compiles the underlying trips, including their origin, destination, and times. Using Google Places, this paper also collected a wide variety of POIs relevant for mobility and clustered them in 10 broad categories. POIs were then associated with each trip origin and destination, which led to the creation of 2-dimensional matrices identifying trip purposes. 

The results of the study are particularly illuminating. They indicate that e-scooters fill specific last-mile needs for certain mobility trips. In particular, the results indicate that e-scooters are primarily used for business and leisure. Business to business trips are prominent for e-scooters, which was a surprise. These trips occur primarily in the afternoon and provide both an affordable and convenient mobility for this population segment. Business from/to parking trips are also significant, with potentially interesting implications for cities and urban planning. Birds are in heavy use for leisure trips to bars and restaurants, especially in the evening and at night (before 9pm). 

The use of e-scooters in connection with transit is small overall and probably due to the relatively high additional cost. However, given the use of e-scooters in conjunction with parking suggests that proper financial incentives  (e.g., joint monthly passes for transit and e-scooters) may change this equation. The lack of e-scooter together with transit is not due to any technical limitation and hence cities may have a significant opportunity to address first/last mile mobility with e-scooter: It requires however the proper incentive structure.

Interestingly, e-scooters exhibit interesting time of day patterns, with a surge in use in the afternoon that is maintained through the evening. This is explained obviously by the trip purposes.

E-scooters are a recent addition to the mobility landscape and it is likely that their use will continue to evolve quickly, given their proliferation in many cities around the world. It would be interesting to perform a similar study in the next few months again to see how ridership is evolving.

\section{Acknowledgements}

This research is partly funded by NSF grant 1854684. Many thanks to Connor Riley for providing us with the raw data. 

\bibliographystyle{plain}
\bibliography{trb_template}

\end{document}